# Basis set convergence and extrapolation of connected triple excitation contributions (T) in computational thermochemistry: the W4-17 benchmark with up to $k$ functions

Jan M.L. Martin

*Department of Molecular Chemistry and Materials Science, Weizmann Institute of Science, 7610001 Reḥovot, Israel. Email: gershom@weizmann.ac.il*

Abstract. The total atomization energy of a molecule is the thermochemical cognate of the heat of formation in the gas phase, its most fundamental thermochemical property. We decompose it into different components and provide a survey of them. It emerges that the connected triple excitations contribution is the third most important one, about an order of magnitude less important than the "big two" contributions (mean-field Hartree-Fock and valence CCSD correlation), but 1-2 orders of magnitude more important than the remainder. For the 200 total atomization energies of small molecules in the W4-17 benchmark, we have investigated the basis set convergence of the connected triple excitations contribution (T). Achieving basis set convergence for the valence triple excitations energy is much easier than for the valence singles and doubles correlation energy. Using reference data obtained from *spdfghi* and *spdfghik* basis sets, we show that extrapolation from quintuple-zeta and sextuple-zeta yields values within about 0.004 kcal/mol RMS. Convergence to within about 0.01 kcal/mol is achievable with quadruple- and quintuple-zeta basis sets, and to within about 0.05 kcal/mol with triple- and quadruple-zeta basis sets. It appears that radial flexibility in the basis set is more important here than adding angular momenta $L$: apparently, replacing *nZaPa* basis sets with truncations of 7ZaPa at $L=n$ gains about one angular momentum for small values of n. We end the article with a brief outlook for the future of accurate electronic structure calculations.

Keywords: coupled cluster theory; triple excitations; basis set convergence; thermochemistry; basis set extrapolation

## 1. Introduction
### 1.1 Electron Correlation and the Correlation Energy

Exact analytical solution of the electronic Schrödinger equation is only possible for one-electron systems; for a "numerically exact" solution (see below), the computational cost scales *factorially* with the number of electrons n.

Already Hartree[1–3] in the 1920s considered a mean-field approximation in which the motions of the electrons are statistically uncorrelated: this gives rise to an approximate n-particle wave function consisting of a product of one-particle wave functions (a "Hartree product"). Fock[4, 5] and Slater[6] simultaneously and independently extended the theory to account for indistinguishable particles: in this "Hartree-Fock theory" (HF), the approximate wavefunction is an antisymmetrized product ("Slater determinant") of one-electron orbitals.

Löwdin[7] defined the difference between the exact *n*-electron energy and the HF energy as the *correlation energy*. (The term itself was first used by Wigner.[8])

$$E_{corr} = E_{exact} - E_{HF} \tag{1}$$

$E_{corr}$ is typically less than one percent of $E_{exact}$. By way of illustration: for one type of systems that has been studied extensively, namely neutral atoms, an asymptotic expansion of the correlation energy in the atomic number Z has the leading terms [9] (in units of 1 Hartree=627.5095 kcal/mol):

$$E_{corr} = -0.02073\ Z \ln Z + 0.0378(9)\ Z\ + \ldots \tag{2}$$

In contrast, the following asymptotic expansion in the atomic number Z holds for the Hartree-Fock energy [10–13]:

$$E_{HF} = -0.7687\ Z^{7/3} + 0.5\ Z^2 - 0.2699\ Z^{5/3} - 0.2240Z + 0.2467\ Z^{2/3} + \ldots \tag{3}$$

While this expression is intended for the large-Z limit, Schwinger[10, 11] observed that eq. (3) is "unreasonably accurate" across the Periodic Table — even at the smallest extreme, hydrogen (Z=1), the error is only 10%.

Combining both expressions, the ratio $E_{corr}/E_{total}$ has the leading term

$$E_{corr}\ /E_{total}= E_{corr}\ /(E_{HF} + E_{corr}) = 0.027 \ln Z\ /\ Z^{4/3} + \ldots \tag{4}$$

It would thus appear, at naïve first sight, that the correlation energy is not very important, as already for Ne atom it has dropped to 0.3% of the total. However, as can be seen in Table 1, nothing could be further from the truth concerning *reaction* energies. Of the dissociation energy of $N_2$ (that is, the bond energy of the N≡N triple bond), for example, Hartree-Fock theory only recovers about *half* — in general, about 80% is the best Hartree-Fock theory can do for bond energies. How is this possible?

The answer lies in two observations. First, as seen in Eq. (3), the energy scale for total energies is huge, several orders of magnitude larger than typical chemical reaction energies: all of computational thermochemistry can be regarded as an exercise in extracting (sometimes very) small differences of very large numbers.

Second, from density functional theory within the local density approximation (LDA), we can consider the dependence of the correlation energy on the electronic density. Perhaps the simplest analytical expression for the LDA correlation energy density is that of Chachiyo[14]:

$$\varepsilon = a \ln(1 + b/r_s + b/r_s^2), \qquad a=(\ln 2 -1)/(2\pi^2),\ b=20.4562557 \tag{5}$$

Where the Wigner-Seitz radius $r_s=(4\pi\rho/3)^{-1/3}$ is the classical average distance between electrons in a homogenous electron gas with density $\rho$. In other words: for low density, the correlation energy is proportional to $\rho^{1/3}$ and for large density to $\ln \rho$.

What thus happens, when one brings together atoms to form a molecule, is that the electron density increases in the bonding region between them, and hence so does the correlation energy.

### 1.2 Coupled Cluster Theory

According to Löwdin's theorem, the exact solution for any n-particle wave function in a given finite basis set can be obtained as the linear combination of the Hartree-Fock reference and all possible excitations out of it, i.e. all possible determinants generated by moving orbitals from occupied to unoccupied (virtual) orbitals. If we group these by excitation levels, we obtain (using intermediate normalization)

$$\Psi_{FCI} = \left(1 + \hat{C}_1 + \hat{C}_2 + \hat{C}_3 + \cdots + \hat{C}_n\right)\Psi_0 \quad (8)$$

Where the single, double,… excitation operators are defined by

$$\hat{C}_1\Psi_0 = \sum_{i,a}^n C_{ia} \, \Psi_{i\to a}; \; \hat{C}_2\Psi_0 = \sum_{i>j,a>b}^n C_{ij,ab} \, \Psi_{ij\to ab}; \, \ldots \qquad (9)$$

where the $i,j,k,$… indices represent occupied orbitals, $a,b,c,$… represent virtual orbitals, and $C_{ia}$, $C_{ij,ab}$, $C_{ijk,abc}$, … are CI coefficients.

The computational cost of FCI scales factorially in both the number of electrons $n$ and the number of basis functions $N$. One could truncate at a given excitation level, which leads to *limited configuration interaction* (or just CI for short): this approach suffers from size extensity errors that may rival actual reaction energies, and hence has become obsolete. Alternatively, one could apply a perturbation theory expression (many-body perturbation theory), which in practice only works well if the gap between occupied and unoccupied orbitals is sufficiently large. In *coupled cluster theory* (see Shavitt and Bartlett for a monograph[15]), the size extensivity problem of CI is eliminated by rewriting the trial wave function as an exponential ansatz:

$$\Psi_{FCI} = \exp\left(\hat{T}\right)\Psi_0 = \exp\left(\hat{T}_1 + \hat{T}_2 + \hat{T}_3 + \cdots + \hat{T}_n\right)\Psi_0 \quad (10)$$

If one does not truncate this expression, one merely has a clumsier way of doing FCI. But if one truncates the cluster operator at a given excitation level, it can be shown (the linked-cluster theorem[16–18]) that the result is rigorously size-extensive.

Consider for example a dimer A…B of two-electron systems A and B at infinite distance. CISD is an exact solution for each monomer on its own, but the dimer wave function at infinite distance will be an antisymmetrized product of the monomer wave functions, which thus includes the cross-term $\psi(A)\,\psi(B)$ . It is clear that the latter entails quadruple excitations with respect to the reference. (This specific kind of quadruple excitations made up of simultaneous and independent double excitations is known as *disconnected* quadruples in coupled cluster lingo.)

In contrast, the CCSD (coupled cluster with all singles and doubles[19]) wavefunction through $\exp(T_2)=T_2+T_2^2/2+T_2^3/6+\ldots$ naturally includes connected quadruple excitations of this type.

To borrow a metaphor from Janesko[20], while a 42-electron full CI wavefunction for benzene is like an impossibly unwieldy "marriage between 42 people…[i]n real life, 42 people in a room don't need to behave like they're all married to each other! 'Manners', simple guidelines for behavior, suffice for most interactions in everyday life." One might see CCSD as one such sets of manners for couples (electron pairs) how to behave with each other on a dance floor. The computational cost of CCSD in a given basis set scales asymptotically as $O(n^2N^4)$, which represents a reduction from exponential to polynomial cost scaling.

CCSD captures the breaking of a single bond quite well: however, for multiple bonds, the next higher term is required. From a somewhat naïve point of view, you might understand this as follows[21]: if you have a low-lying singly excited determinant, then it will be important in the wavefunction, but so will double excitations from it, i.e., connected triple excitations w.r.t. the reference determinant. If you have a low-lying doubly excited determinant, you will have a substantial term of doubles out of doubles, i.e., connected quadruples.

Unfortunately, CCSDT[22, 23] and CCSDTQ[24] have computational costs that scale asymptotically as $O(n^3N^5)$ and asymptotically as $O(n^4N^6)$, respectively. A felicitous compromise between accuracy and computational cost is represented by the CCSD(T) method[25, 26], where CCSD is augmented by a quasiperturbative estimate of the contribution of connected triples, $T_3$, with an asymptotic cost scaling $O(n^3N^4)$ rather than $O(n^3N^5)$.

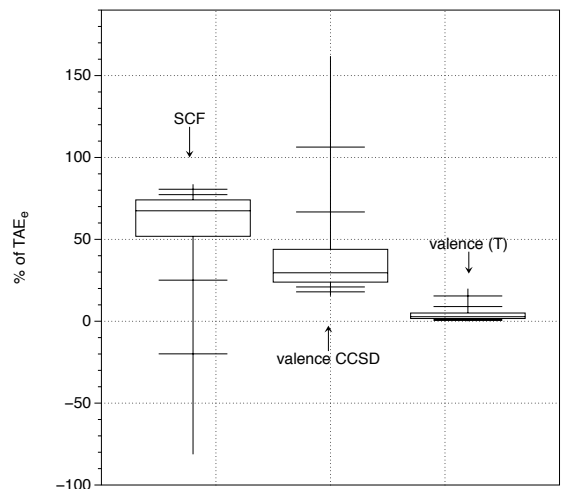

**Figure 1. Box-and-whiskers plot of the total atomization energy contributions in the W4-17 dataset[27] of 200 small molecules. Outer fences encompass 95% of the set, inner fences 80%, boxes 50%. Vertical lines span from population minimum to maximum.**

Empirically[21, 27, 28], and heavily relying on the arbitrary-order coupled cluster code developed by Kállay and coworkers[29–31], we have found that CCSD(T) represents a felicitous error compensation between neglect of higher-order triples (which are generally antibonding) and of connected quadruple excitations (which are universally bonding):

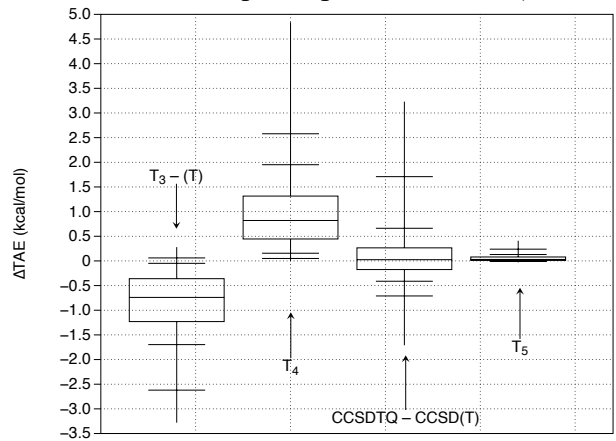

**Figure 2. Box-and-whiskers plot of the total atomization energy contributions of higher-order corrections in the W4-17 dataset[27]. Outer fences encompass 95% of the set, inner fences 80%, boxes 50%. Vertical lines span from population minimum to maximum.**

Stanton[32] gives a theoretical rationale for this in terms of perturbation theory with Löwdin partitioning[33] starting from CCSD as the zero order reference; see also Kállay and Gauss[34]. Suffice to say that CCSD(T) has become the "gold standard" (T. H. Dunning[35]) of wavefunction ab initio theory.

### 1.3 Broader Context of the Problem

Aside from some niche and emerging methods, computational quantum chemistry today is synonymous with two primary approaches. In wavefunction theory (WFT), highly accurate approximate solutions to the Schrödinger equation can be systematically refined to accuracy levels competing with the best experiments — or indeed surpassing them. The price one pays for this is their very steep CPU time scaling with the size of the system (N): for a truly exact solution within the given basis set, like full CI, this would be factorial, while for the "gold standard" CCSD(T) method, it is 'merely' $O(N^7)$.

In contrast, density functional theory (DFT) has comparatively gentle system size scaling, $O(N^3)$ or gentler — at the expense of introducing an unknown, and perhaps unknowable, exchange-correlation (XC) functional. Over the years, DFT has established itself as the 'bread and butter tool' of the molecular modeling community. But, while exchange-correlation functionals have come a long way, they still have to go quite a distance before they become competitive in accuracy with high-level WFT approaches.

Recent developments in localized orbital approaches, such as the DLPNO-CCSD(T) method of Neese and coworkers[36, 37], PNO-LCCSD(T) by Werner and coworkers[38], and LNO-CCSD(T) by Nágy and Kállay[39], appear to have revived interest in WFT methods even among computational chemists with modest computational resources. For sufficiently large systems, these methods boast near-linear scaling, at the expense of introducing numerous cutoffs and threshold that arguably introduce an empiricism of precision (instead of the empiricism of accuracy inherent in DFT methods).

Another approach for large systems that has been gaining ground is to use Δ-machine learning[40] to correct inexpensive calculated values to near-WFT quality. All such approaches require a substantial amount of high-accuracy WFT data for small molecules as a 'training set'. The advantages in using accurate WFT data as a 'primary standard'[1] instead of experimental data are manifold. First of all, one is not restricted to the parts of chemical space for which experimental data of the required accuracy is available. Second, data do not need to be isolated from experimental 'confounding factors' to bring them 'on the same page' with the simulation. All practical WFT approaches today rely on finite basis sets. This means that establishing basis set convergence becomes an essential aspect of any WFT study.

Of all the ground-state properties of an atom or molecule, the most fundamental is the total energy. However, with the possible exception of hydrogen atom, absolute energies are 2-3 orders of magnitude larger than any reaction energy of chemical interest (cf. eq. 3). Consider that for just bare neon atom, Z=10, we are already talking about more than 80,400 kcal/mol! These numbers only become more staggering as we add more atoms: clearly attempting to reproduce total energies to within, say, 1 kcal/mol is a Sisyphean exercise.

---

[1] The term originates in quantitative analytical chemistry and is used here by analogy.

The next step down would be total atomization energies (TAEs), i.e., the energy required to separate a neutral molecule into its constituent atoms. (For charged molecules, this can be combined with ionization potentials and electron affinities.) The computed total atomization energy, in combination with atomic heats of formation in the gas phase, can be directly related to the gas-phase heat of formation. (Thanks to the Active Thermochemical Tables project[41–43], a consistent set of mixed experimental-theoretical values based on a global thermochemical network is available.)

### 1.4 Decomposition of the Total Atomization Energy

The total atomization energy of a molecule $A_mB_n\ldots$ corresponds to the sum of all the bond energies in the molecule. It is defined as the energy required to dissociate the molecule into separate atoms in their electronic ground states, all in the gas phase:

$$TAE_e[A_mB_n\ldots] = m\,E[A] + n\,E[B] + \ldots - E[A_mB_n\ldots] \qquad (11a)$$
$$TAE_0[A_mB_n\ldots] = TAE_e[A_mB_n\ldots] - ZPVE[A_mB_n\ldots]. \qquad (11b)$$

where $TAE_e$ is the total atomization energy in the hypothetical motionless state, $TAE_0$ that at 0 kelvin, and ZPVE is the zero-point vibrational energy[44, 45] of the molecule. $TAE_0$ is the thermochemical cognate of the gas-phase heat of formation at absolute zero:

$$\Delta H^\circ_{f,0}[A_mB_n\ldots] = m\,\Delta H^\circ_{f,0}[A(g)] + n\,\Delta H^\circ_{f,0}[B(g)] + \ldots - TAE[A_mB_n\ldots] \quad (12)$$

As discussed in detail in, e.g., Refs.[27, 46, 47], the electronic components of the total energy (and hence also of the $TAE_e$) of a small row 1 or 2 molecule can be decomposed into the following components (see Table 1 for some representative molecules from the 200-molecule W4-17 dataset[27]):

- The Hartree-Fock SCF component
- The valence CCSD (coupled cluster with all singles and doubles[19]) correlation energy component
- The valence (T) connected triples[26, 48] component, which corresponds to the CCSD(T) – CCSD difference
- valence post-CCSD(T) correlation effects (discussed in detail in Refs.[21, 49])
- the contribution of inner-shell correlation (discussed and reviewed in detail in Ref.[50])
- scalar relativistic corrections[51]
- spin-orbit coupling (which for light closed-shell species effectively amounts to the fine structures of the constituent atoms[52])
- diagonal Born-Oppenheimer corrections (DBOC)[53]

All computational protocols for high-accuracy computational thermochemistry include all these terms in one fashion or another — be it Weizmann-4 (W4) and W4-F12 from our own group[21, 46, 54], HEAT (High-accuracy Extrapolated Ab initio Thermochemistry) developed by a multinational team centered around Stanton[47, 55–57], or the Feller-Peterson-Dixon (FPD) approach[58–64].

As expected, and as seen in Table 1, Hartree-Fock and valence CCSD correlation are the two dominant contributions: inner-shell correlation is two orders of magnitude below that, as most of it cancels between the molecule and its separated atoms. (For a detailed discussion of its basis set convergence, see Ref.[50] and references therein.) Valence connected triple excitations, (T), are an order of magnitude less important than valence CCSD, but still outweigh all remaining components by 1-2 orders of magnitude. The latter is true even for $O_3$ and $N_2O_4$ where nondynamical correlation effects[65] drive post-CCSD(T) contributions into the kcal/mol range — it was actually shown a decade and a half ago[54] that the relative importance of (T) in the CCSD(T) TAE is a good predictor for the thermochemical importance of post-CCSD(T) correlation effects.[2]

**Table 1: decomposition of the total atomization energies (kcal/mol) in the hypothetical motionless state of six representative species from the W4-17 dataset. All data are taken from the Supporting Information of Ref.[27]; see the reference for computational details.**

|                        | $N_2O_4$ | $C_6H_6$ | $SiF_4$ | $SO_3$ | $O_3$ | $C_3H_8$ |
|------------------------|----------|----------|---------|--------|--------|----------|
| Hartree-Fock           | 112.87   | 1045.01  | 448.41  | 159.72 | -45.09 | 785.34   |
| Valence CCSD           | 313.27   | 290.71   | 119.07  | 165.45 | 163.94 | 209.05   |
| Valence (T)            | 42.85    | 26.70    | 10.12   | 20.28  | 25.62  | 10.12    |
| CCSDT – CCSD(T)        | -2.92    | -2.62    | -1.14   | -1.68  | -1.34  | -0.63    |
| CCSDTQ – CCSDT         | 4.21     | 1.63     | 0.46    | 1.75   | 3.81   | 0.38     |
| Post-CCSDTQ            | —        | —        | 0.00    | 0.21   | 0.41   | 0.02     |
| Inner-shell correlation| 1.40     | 7.37     | 0.84    | 1.16   | 0.08   | 3.61     |
| Scalar relativistics   | -1.00    | -0.99    | -1.90   | -1.85  | -0.25  | -0.58    |
| Spin-orbit coupling    | -0.89    | -0.51    | -1.97   | -1.23  | -0.67  | -0.25    |
| DBOC                   | 0.06     | 0.23     | 0.05    | 0.05   | -0.03  | 0.12     |

That leaves us with HF, valence CCSD, and valence (T) as the "big three". Now for small molecules, the HF contribution is amenable to numerically exact calculation[66–69], or to calculations with very large finite basis sets, to the point that they become de facto exact[70–72]). The valence CCSD contribution with its slow basis set convergence continues to be the focus of much research using basis set extrapolation[54, 73–77] (*vide infra*) and explicitly correlated approaches.[46, 78–80] Thus (T) is left as the remaining major contributor, on which we will focus in the present chapter.

### 1.5 Gaussian Basis Sets

Orbital-based electron correlation methods in practice require a finite basis set, as otherwise the Löwdin expansion cannot be finite.

In computational solid-state physics, plane waves form a very natural basis set, where only an energy cutoff needs to be specified. Such basis sets by construction assume a periodic (as in: unit cell-based) system, and hence are in practice unsuitable for molecular calculations. The latter are almost universally carried out in basis sets that are a linear combination of functions that at least *resemble* atomic orbitals. In the DFT world, both numerical orbitals (e.g., in FHI-AIMS[81, 82]) and Slater-type orbitals (e.g., in ADF[83, 84]) are used in some codes, but by far the most commonly used are Gaussian type orbitals (GTOs), i.e., products of spherical harmonics $Y(\theta,\phi)$ with a Gaussian function of $r$: A $Y(\theta,\phi).\exp(-\zeta r^2)$. The main reason for their near-universal adoption is computational convenience, i.e., the Gaussian product theorem, which dramatically speeds up evaluation of the four-center two-electron integrals that occur in WFT calculations.

---

[2] We also note in passing that as one goes further down the periodic table, relativistic effects will *eventually* come to rival the major contributors[171].

Gaussian basis sets have been extensively reviewed, e.g., by Davidson and Feller[85], Shavitt[86], Peterson[87], Hill[88], Jensen[89], and most recently by Nagy and Jensen[90], who also cover other basis set types. For correlated wavefunction calculations, the correlation consistent polarized n-tuple zeta family[87, 91] in its many variants has become something of a de facto standard. Atomic natural orbitals[92–94] have recently been revisited[95, 96]; polarization consistent basis sets[97, 98] have seen some adoption in DFT, while the Karlsruhe basis sets[99] offer something of a compromise between the demands of WFT and DFT. Very recently, the nZaPa sets of Petersson[73, 100] offer a numerically somewhat better-behaved alternative to the correlation consistent family.

Recently, explicitly correlated methods[78, 79] in which terms explicitly dependent on interelectronic distances such as F12 geminals[101] are added to the basis set, have become a powerful addition to the WFT toolbox due to the greatly accelerated basis set convergence in MP2 and CCSD. Unfortunately for the subject at hand, triple excitations do not benefit from F12, as was shown in great detail by Köhn[102, 103].

### 1.6 Basis Set Extrapolation

The convergence of the correlation energy is quite slow, but asymptotically systematic. For the MP2 (second order Møller-Plesset[104] perturbation theory) correlation energy of helium-like atoms, Schwartz[105, 106] showed that the "partial wave increment" of angular momentum $\ell$ — that is, the total contribution to the correlation energy of all basis functions with angular momentum $\ell$ — will be of the form:

$$E^{(2)}(\ell) = A\ (\ell +1/2)^{-4} + B/(\ell +1/2)^{-6} + O(\ell^{-8}) \tag{13}$$

Hill[107] generalized this result to variational energies as

$$E^{(CI)}_{(l)}(\ell) = A\ (\ell +1/2)^{-4} + C/(\ell +1/2)^{-6} + O(\ell^{-6}) \tag{14}$$

In an analytical tour de force, Kutzelnigg and Morgan[108] generalized this work to arbitrary pair correlation energies in an atom: they found that for same-spin correlation energies, the expansion starts at $(\ell+1/2)^{-6}$ rather than $(\ell+1/2)^{-4}$.

In order to estimate residual basis set incompleteness for a basis set that saturates partial waves through angular momentum L, we need to sum contribution from L= $\ell$ +1 through infinity. This can be done analytically with the help of the polygamma[109] function $\psi^{(n)}(x)$. Replacing the latter by their asymptotic series expansions, we finally obtain as the leading term

$$E^{(2)}(L) = E^{(2)}_\infty + A.L^{-3} + O(L^{-5}) \tag{15}$$

From a second perspective, the principal expansion[110], Bunge[111] and Carroll, Silverstone, and Metzger (CSM) [112] independently found that the contribution of a single atomic orbital with quantum numbers n,$\ell$,m is essentially independent of the angular quantum number $l$ and the magnetic quantum number $m$, and depends on the principal quantum number $n$ as

$$\delta E_{n,\ell,m} = -A(n-1/2)^{-6} \tag{16}$$

For a given $n$, $\ell$ can run from 0 to $n-1$, and $m$ in turn from $-\ell$ to $\ell$. Thus, there are $\sum_{l=0}^{n-1}(2\ell + 1) = n^2$ essentially equal contributions. The contribution for each principal quantum number n thus acquires a $\propto n^{-4}$ leading dependence, and as above, summation leads to a leading $\propto n^{-3}$ dependence of the overall basis set incompleteness. Indeed, this "principal expansion" structure is exhibited by both of the major types of basis set sequences for correlated WFT calculations: the correlation consistent basis sets[87] (through clustering by atomic correlation energy contributions) and the atomic natural orbital basis sets[92–94] (through clustering by natural orbital occupation number).

If the correlation energy (or a contribution to it) converges proportionally to $L^{-\beta}$, then the basis set limit is trivially obtained from energies with two successive L as

$$E = E[L] + (E[L] - E[L-1])/((L/L-1)^\beta - 1) \tag{17}$$

That an expression for pair correlation energies might also apply to the complete atomic correlation energy perhaps seems at least plausible. (Petersson[113] considered separate extrapolations of pair energies, using an L-shift as an adjustable parameter, *vide infra*.) That the correlation energy of a molecule, however, would behave similarly to a spherical requires more of a 'leap of faith': fortunately, during initial explorations for thermochemistry[114, 115] we found that this is basically the case. This implies, incidentally, that the molecular correlation energy behaves largely like 'atoms in molecules'. A very simple formula due to Halkier et al.[75] is widely used:

$$E(L) = E_\infty + A/L^3 \quad \text{or hence } E_\infty = E(L) + [E(L) - E(L-1)]/[(L/L-1)^3 - 1] \tag{18}$$

It was soon discovered empirically that better results could be obtained for smaller basis sets, and for specific components of the correlation energy, if the exponents were used as adjustable parameters: this, as well as the Petersson approach[73, 116] of using L-shifts in the same way, is actually equivalent to Schwenke's [74] twopoint linear extrapolation, as discussed in detail here[117].

(Extrapolation of the total energy for smaller basis sets leads to the misleading conclusion that overall convergence is exponential[118], which works well enough for HF and DFT energies but causes serious underestimates of the WFT correlation energy.)

While it has been understood for about two decades (e.g., Helgaker and coworkers[119]) that (T) converges faster with the basis set than CCSD, this is one aspect where the Karton-Martin W4-11 and W4-17 thermochemical benchmarks[27, 28] might benefit from more accurate calibration. As a by-product thereof, we will present revised extrapolations of the (T) correlation energy for common basis set pairs.

Two-point basis set extrapolations, inspired by the Schwartz-Kutzelnigg partial-wave expansion[106, 107, 120] and the Klopper-Helgaker principal expansion[110], take the form:

$$E(L) = E_\infty + A.(L + a)^{-\alpha} \tag{19}$$

From which follows that

$$E_\infty = E(L) + \frac{E(L) - E(L-1)}{\left(\frac{L+a}{L+a-1}\right)^\alpha - 1} \tag{20}$$

where L is the largest angular momentum in the basis set used for calculating the total energy E(L), and the exponent $\alpha$ and the L shift $a$ are extrapolation parameters specific to the level of theory. Typically, one or both of $\alpha$ and $a$ are frozen: Halkier et al.[75] set $a$=0 and $\alpha$=3, the present author[114, 115] originally favored fixed $a$=0 or $a$=1/2 and fitted $\alpha$, while the Petersson group[73] favor fixed $\alpha$=3 and fitted $a$. (Klopper[121], building on the landmark analytical work of Kutzelnigg and Morgan[122], advocated separate $L^{-3}$ and $L^{-5}$ extrapolation of singlet- and triplet-coupled CCSD pair correlation energies, respectively, which is the approach we adopted in W4 theory[54].) Schwenke[74] instead proposed to simply consider a two-point linear extrapolation of the form:

$$E_\infty = E(L) + A_{L-1,L}[E(L) - E(L-1)] \tag{21}$$

where $A_L$ is a coefficient specific to the basis set pair and the level of theory.[3] As discussed in Ref.[123], the Schwenke form is mathematically equivalent to eq. (2) if the following relationships apply

$$E_L = E_\infty + \frac{B}{L^\alpha} \quad \text{if} \ \ \alpha = \frac{log\left(1 + \frac{1}{A_{L-1,L}}\right)}{log\left(\frac{L}{L-1}\right)} \tag{22}$$

$$E_L = E_\infty + \frac{D}{(L+a)^3} \quad \text{if} \ \ a = \frac{1}{\left(1 + \frac{1}{A_{L-1,L}}\right)^{1/3} - 1} + 1 - L \tag{23}$$

Or, conversely

$$A_{L-1,L} = \frac{1}{\left(\frac{L+a}{L-1+a}\right)^\alpha - 1} \tag{24}$$

Ranasinghe and Petersson (RP)[73] determined Schwenke coefficients for (T) and their nZaPa basis set family[73] (n=2–7, the largest basis set topping out at $k$ functions) by fitting to a fairly large set of total energies of small first-and second-row species. The reference data for the MP2, CCSD-MP2, and (T) components were obtained by least-square fitting of each component *individually* to expressions of the form $E_\infty$ +A(L+$a$)$^{-3}$ analogous to the CBS pair extrapolation by Petersson and coworkers[113], using $E_\infty$ and $a$ as fit parameters *for each system* separately to nZaPa (n=4,5,6,7). Next, they fitted two-point Schwenke coefficients $A_L$ for {L-1,L} pairs for each component: specifically, for the (T) component, they found $A_{2,3}$=0.466, $A_{3,4}$=0.600, $A_{4,5}$=0.849, $A_{5,6}$=1.164, and $A_{6,7}$=1.580. They noted that these coefficients are reproduced fairly well by the extrapolation formula:

$$E(L) = E_\infty + A. \left[ (L - \frac{2}{3})^{-3} - \frac{7}{8}(L - \frac{2}{3})^{-5} \right] \tag{25}$$

---

[3] He recommends eschewing nonlinear 3-point formulas, as they are not size-consistent. We note in passing that Schwenke also presents separate extrapolation coefficients for the Klopper-style[121] singlet-coupled and triplet-coupled CCSD correlation energy components.

which yields $A_{2,3}$=0.446, $A_{3,4}$=0.604, $A_{4,5}$=0.891, $A_{5,6}$=1.199, and $A_{6,7}$=1.517. For comparison, a simple $L^{-3}$ extrapolation would yield, respectively, 0.421, 0.730, 1.049, 1.374, and 1.701.

In this chapter, we will consider basis set convergence of (T) for the W4-17 atomization energies in detail, and, as a by-product, obtain extrapolation parameters for a number of basis sets where they were hitherto unavailable. We will show that, leaving aside the behavior of contributions to the absolute correlation energy, TAE[(T)] actually converges reasonably rapidly with the basis set and can be obtained to 0.01 kcal/mol accuracy using no more than quintuple-zeta basis sets. We will also present evidence that radial flexibility of the basis set is more important for (T) than 'piling on' higher angular momenta.

## 2. Computational methods

All electronic structure calculations with basis set sequences requiring at most $i$ functions were carried out using the MOLPRO 2020.2 electronic structure package[124] running on the ChemFarm HPC facility of the Faculty of Chemistry at the Weizmann Institute of Science. The 7ZaPa basis set of RP[73] requires $k$ functions, which exceed the supported maximum angular momentum of MOLPRO, and hence Gaussian 16 rev. C.01 was employed for these[125]. Reference geometries of the W4-17 dataset were used 'as is' from the supporting information of Ref.[27] For open-shell cases, the restricted open-shell CCSD(T) definition of Watts et al.[26] was used throughout. Basis sets were used from the internal library of MOLPRO, except for the nZaPa, which were downloaded from https://www.basissetexchange.org, version 2 of the Basis Set Exchange [126].

In the correlation consistent[87, 91] basis set sequences, we used cc-pVnZ on hydrogen[91, 127], aug-cc-pVnZ on first-row elements[128], and aug-cc-pV(n+d)Z on second-row elements[129, 130]. (The addition of a high-exponent d function is required for second-row elements in high oxidation states to ensure proper description of the 3d orbital[130–132], which acts as a back-bonding recipient[133] from O and F.) This combo is indicated below as haVnZ+d. We also considered correlation core-valence basis sets[134, 135] aug-cc-pCVnZ (or ACVnZ for short); in part, total *valence* energies from Ref.[50] were recycled for this purpose.

In terms of basis sets for explicit correlation, we considered the cc-pVnZ-F12 basis sets[136, 137] as well as their augmented counterparts,[138] both in conventional CCSD(T) and in CCSD(T)-F12b[139] contexts. In the latter, the auxiliary basis sets employed were MOLPRO's defaults for exchange[140], RI-MP2[141], and complementary auxiliary basis sets[142], with the recommended geminal exponents for each basis sets.

For the W4-08 subset of W4-17, we were able to carry out CCSD(T)/7ZaPa calculations, except for NCCN (dicyanogen) which diverged for numerical reasons. CCSD(T)/haV6Z+d calculations proved possible for all of W4-17 except for benzene (near-linear dependence in the basis set) and n-pentane (lack of scratch storage); all of W4-17 could be treated with the remaining basis sets.

We will use the {n-1,n} notation for extrapolation throughout: for example, cc-pV{5,6}Z denotes basis set extrapolation from cc-pV5Z and cc-pV6Z basis sets.

### 3. Results and discussion

#### 3.1 Convergence in a model system: neon atom

How does (T) really converge for large basis sets, and can we make any sense of the somewhat erratic behavior for small basis sets? Thanks to the work of Barnes, Petersson, and coworkers[100, 116], we have nZaPa basis sets available for neon up to n=10 (!). In addition, from Feller, Peterson, and Crawford[143], we have cc-pVnZ basis sets (and data) for the same system up to n=10.

All relevant data are given in Table 2. We were able to reproduce the data from Refs.[116, 143] (and hence also to extract individual components); In addition, we truncated the 10ZaPa and cc-pV10Z basis sets at successive angular momenta L=2–9, leading to the results labeled "partial wave" in Table 2.

Barnes et al.[116] already pointed out that $E_4(T)$ will be dominated by the $E_{4aab}$ and $E_{4bba}$ mixed-spin components, rather than the same-spin $E_{4aaa}$ and $E_{4bbb}$ terms. But how does this look in practical terms? In the upper pane of Table 2, one can see the $E_{4aaa}/E_{4aab}$ ratio for neon atom as a function of maximum angular momentum L≤9 for both the cc-pVnZ and nZaP basis set sequences.

**Table 2: basis set convergence of the (T) correlation energy (a.u.) for neon atom**

| L | $1000E_{4(T)aaa}/E_{2aa}$ | | $1000E_{4(T)aab}/E_{2ab}$ | | $1000E_{4(T)aaa}/E_{2aa}$ | | $1000E_{4(T)aab}/E_{2ab}$ | | $E_{4aaa}/E_{4aab}$ | | $E_{4aaa}/E_{4aab}$ | |
|---|---|---|---|---|---|---|---|---|---|---|---|---|
| | Partial Wave | VnZ | Partial Wave | VnZ | Partial Wave | nZaP | Partial Wave | nZaP | Partial Wave | VnZ | Partial Wave | nZaP |
| 2 | 3.94 | 1.08 | 12.94 | 4.00 | 3.94 | 1.19 | 12.93 | 4.03 | 0.0558 | 0.0509 | 0.0558 | 0.0555 |
| 3 | 4.43 | 3.26 | 13.47 | 10.71 | 4.43 | 3.33 | 13.46 | 10.74 | 0.0559 | 0.0541 | 0.0559 | 0.0550 |
| 4 | 4.50 | 4.12 | 13.28 | 12.35 | 4.50 | 4.14 | 13.27 | 12.39 | 0.0551 | 0.0560 | 0.0550 | 0.0559 |
| 5 | 4.51 | 4.39 | 13.20 | 12.95 | 4.50 | 4.38 | 13.19 | 12.92 | 0.0545 | 0.0552 | 0.0545 | 0.0551 |
| 6 | 4.50 | 4.46 | 13.14 | 13.07 | 4.50 | 4.45 | 13.13 | 13.04 | 0.0543 | 0.0547 | 0.0542 | 0.0546 |
| 7 | 4.50 | 4.48 | 13.11 | 13.08 | 4.50 | 4.48 | 13.10 | 13.07 | 0.0541 | 0.0544 | 0.0541 | 0.0544 |
| 8 | 4.50 | 4.49 | 13.09 | 13.08 | 4.50 | 4.49 | 13.08 | 13.08 | 0.0541 | 0.0542 | 0.0541 | 0.0542 |
| 9 | | 4.50 | | 13.08 | | 4.50 | | 13.07 | | 0.0541 | | 0.0541 |

| | E(T) | E(T) | EcorrCCSD | EcorrCCSD | E(T) | E(T) | EcorrCCSD | EcorrCCSD |
|---|---|---|---|---|---|---|---|---|
| 2 | -0.004915 | -0.000981 | -0.258883 | -0.189017 | -0.004912 | -0.001044 | -0.259024 | -0.195672 |
| 3 | -0.006004 | -0.004259 | -0.293961 | -0.266347 | -0.006001 | -0.004245 | -0.294119 | -0.268254 |
| 4 | -0.006291 | -0.005575 | -0.306758 | -0.294682 | -0.006288 | -0.005538 | -0.306912 | -0.295736 |
| 5 | -0.006394 | -0.006098 | -0.310965 | -0.305489 | -0.006391 | -0.006099 | -0.311117 | -0.305850 |
| 6 | -0.006433 | -0.006290 | -0.312767 | -0.309906 | -0.006430 | -0.006295 | -0.312919 | -0.310129 |
| 7 | -0.006450 | -0.006379 | -0.313642 | -0.312111 | -0.006448 | -0.006377 | -0.313793 | -0.312242 |
| 8 | -0.006458 | -0.006425 | -0.314089 | -0.313327 | -0.006455 | -0.006421 | -0.314241 | -0.313404 |
| 9 | | -0.006449 | | -0.313981 | | -0.006446 | | -0.314103 |

"Partial wave" corresponds to cc-pV10Z or 10ZaP truncated at angular momentum L

One observation we can make is how stable the $E_4(T)_{aaa}/E_4(T)_{aab}$ ratio remains as a function of basis set, holding steady at about 0.054 except for the smallest cc-pVnZ and nZaP basis sets. Another is that the $E_{4(T)aaa}/E_{2aa}$ ratio for sufficiently large n stabilizes at 0.0045, and the $E_{4(T)aab}/E_{2ab}$ ratio at 0.0131. However, for small *n* these ratios are much smaller, 0.001 for cc-pVDZ and 0.004 for 2ZaP. In contrast, the partial-wave series yield ratios that are close to the limiting values even for L=2, i.e., cc-pV10Z-$L_{max}$=2 and 10ZaPa-$L_{max}$=2.

Comparison of the (T) correlation energies (Table 2, lower pane) also shows that for L=2 and L=3, the partial-wave expansions recover a dramatically larger part of the (T) limit than the principal expansions. In general, the basis set convergence of the (T) energy is somewhat erratic for the cc-pVnZ and nZaP basis sets for small n, and much smoother for the partial wave expansions. This is less pronouncedly the case for the CCSD correlation energies, which may indicate that radial flexibility is more of an issue for the (T) term than for the MP2 or CCSD correlation energy.

### 3.2 Convergence and basis set extrapolation for W4-17

Let us now turn to the W4-17 dataset (largest molecule: benzene) and its subset[144] W4-08 (largest molecules: $B_2H_6$ and $C_2H_6$). Root mean square deviations (RMSDs) and fitted extrapolation parameters are presented in Table 3. Full source data are made available as an Excel workbook in the Electronic Supporting Information.

**Table 3: RMSD (kcal/mol) from best reference data (indicated as REF) for the W4-17 dataset and its W4-08 subset, as well as fitted extrapolation exponents (and equivalent Schwenke coefficients and Petersson shifts)**

| | RMSD for W4-08 | Pseudo-Marchetti-Werner[a] | Simple $L^{-3}$ | Petersson, Schwenke, or Hill[b] | Fitted to W4-08 | Optimized Schwenke coefficient[c] $A_{\{L-1,L\}}$ | Equiv. exponent[d] $\alpha$ | Equiv. L shift[e] $a$ | W4-17 Fitted to W4-08 |
|---|---|---|---|---|---|---|---|---|---|
| | Largest basis set | | W4-08 Extrapolated | | | | | | |
| {2,3}ZaPa | 0.812 | | 0.231 | 0.287 | 0.207 | 0.372 | 3.219 | -0.165 | 0.204 |
| {3,4}ZaPa | 0.329 | 0.123 | 0.040 | 0.048 | 0.030 | 0.676 | 3.156 | -0.170 | 0.029 |
| {4,5}ZaPa | 0.147 | 0.062 | 0.043 | 0.013 | 0.010 | 0.803 | 3.626 | -0.770 | 0.007 |
| {5,6}ZaPa | 0.076 | 0.037 | 0.018 | 0.008 | 0.005 | 1.077 | 3.601 | -0.913 | 0.004 |
| {6,7}ZaPa | 0.047 | 0.025 | 0.004 | REF | 0.000 | 1.580 | 3.181 | -0.368 | N/A |
| 7ZaPa-L_max={2,3} | 0.302 | | | | 0.078 | 0.636 | 2.331 | 0.700 | |
| 7ZaPa-L_max={3,4} | 0.141 | | | | 0.016 | 0.873 | 2.654 | 0.450 | |
| 7ZaPa-L_max={4,5} | 0.072 | | | | 0.009 | 1.226 | 2.673 | 0.546 | |
| haV{D,T}Z+d | 0.731 | | 0.199 | 0.190 | 0.188 | 0.385 | 3.155 | -0.120 | 0.176 |
| haV{T,Q}Z+d | 0.305 | | 0.040[g] | 0.040 | 0.039 | 0.708 | 3.062 | -0.070 | 0.059 |
| haV{Q,5}Z+d | 0.135 | | 0.044[g] | 0.009 | 0.009 | 0.794 | 3.654 | -0.799 | 0.010 |
| haV{5,6}Z+d | 0.073 | | 0.013[g] | 0.006 | 0.004 | 1.180 | 3.367 | -0.596 | REF |
| ACV{D,T}Z | 0.707 | | 0.242 | | 0.150 | 0.324 | 3.473 | -0.331 | 0.130 |
| ACV{T,Q}Z | 0.260 | | 0.040 | | 0.032 | 0.666 | 3.186 | -0.201 | 0.034 |
| ACV{Q,5}Z | 0.117 | | 0.034 | | 0.008 | 0.815 | 3.587 | -0.730 | 0.006 |
| ACV{5,6}Z | 0.063 | | 0.012 | | 0.004 | 1.155 | 3.421 | -0.672 | 0.003 |
| V{D,T}Z+d | 0.984 | | 0.246 | 0.304 | 0.246 | 0.423 | 2.991 | 0.007 | 0.270 |
| V{T,Q}Z+d | 0.422 | | 0.046 | 0.054 | 0.046 | 0.746 | 2.957 | 0.050 | 0.062 |
| V{Q,5}Z+d | 0.188 | | 0.060 | 0.020 | 0.015 | 0.800 | 3.633 | -0.777 | 0.017 |
| V{5,6}Z+d | 0.097 | | 0.029 | 0.007 | 0.006 | 1.062 | 3.639 | -0.960 | 0.010 |
| def2-{SVP,TZVPP}[f] | 1.010 | | 0.238 | N/A | 0.238 | 0.424 | 2.989 | 0.009 | 0.240 |
| def2-{TZVPP,QZVP}[f] | 0.411 | | 0.060 | N/A | 0.051 | 0.680 | 3.144 | -0.158 | 0.071 |
| V{D,T}Z-F12 orb. | 0.687 | | | N/A | 0.107 | 0.672 | 2.247 | 0.817 | 0.101 |
| V{T,Q}Z-F12 orb. | 0.299 | | | N/A | 0.031 | 0.819 | 2.774 | 0.281 | 0.036 |
| V{Q,5}Z-F12 orb. | 0.161 | | | N/A | 0.026 | 1.050 | 2.998 | 0.002 | 0.039 |
| V{D,T}Z-F12 F12b | 0.679 | | | 0.206 | 0.119 | 0.708 | 2.172 | 0.931 | 0.113 |
| V{T,Q}Z-F12 F12b | 0.316 | | | 0.051 | 0.038 | 0.861 | 2.679 | 0.414 | 0.045 |
| V{Q,5}Z-F12 F12b | 0.169 | | | N/A | 0.035 | 1.119 | 2.861 | 0.217 | 0.050 |
| aV{D,T}Z-F12 orb. | 0.503 | | | N/A | 0.105 | 0.668 | 2.256 | 0.804 | |
| aV{T,Q}Z-F12 orb. | 0.227 | | | N/A | 0.024 | 0.818 | 2.775 | 0.280 | |
| aV{Q,5}Z-F12 orb. | 0.121 | | | N/A | 0.024 | 1.092 | 2.913 | 0.134 | |

(a) Eq. (27)

(b) With extrapolation coefficients taken from RP[73] for nZaPa, Schwenke[74] for AVnZ, and Ref.[145, 146] for VnZ-F12

(c) E(CBS)= $E(L) + A_{\{L-1,L\}}\, [E(L) - E(L-1)]$

(d) E(CBS)= $E(L) + [E(L) - E(L-1)]/[(L/L-1)^{\alpha} - 1]$

(e) E(CBS)= $E(L) + [E(L) - E(L-1)]/[((L+a)/(L+a-1))^3 - 1]$

(f) def2-{TZVPP,QZVP} extrapolation coefficient for valence CCSD, obtained in the same fashion, is $A_{3,4}$=0.715, or $\alpha$=3.041, or a=0.046. For valence MP2, $A_{3,4}$=0.896, or $\alpha$=2.605, or a=0.524.

(g) 3-point $E_\infty + A.L^{-3} + B.L^{-5}$ extrapolation: haV{D,T,Q}Z 0.055 kcal/mol; haV{T,Q,5}Z 0.062 kcal/mol; haV{Q,5,6}Z 0.009 kcal/mol.

The reference data for W4-08 were obtained by {6,7}ZaPa extrapolation using RP's optimized $A_{6,7}$=1.580. However, the extrapolation covers just 0.047 kcal/mol RMSD from the raw 7ZaPa numbers, and is reasonably insensitive to the precise value of $A_{6,7}$: substituting 1.701 (which corresponds to simple $L^{-3}$ extrapolation) changes the values by just 0.004 kcal/mol RMS. This means our reference data are not an artifact of details of the extrapolation — at least not to any energetic resolution we can realistically hope to achieve.

Indeed, all three {h,i} extrapolation options — V{5,6}Z+d, A'V{5,6}Z+d, and {5,6}ZaPa — have RMSDs below 0.01 kcal/mol from the reference — regardless of whether one uses extrapolation coefficients from Schwenke and RP, or those optimized in the present work against the {6,7}ZaPa reference data. Using the simple $L^{-3}$ extrapolation causes somewhat larger errors, especially for V{5,6}Z+d. The bottom line, however: it *is* possible to achieve 0.01 kcal/mol accuracy in the (T) term with 'just' *spdfgh* and *spdfghi* (i.e., L={5,6}) basis sets.

What about dialing down both basis sets one step? Here 0.01 kcal/mol is possible with both {4,5}ZaPa and A'V{Q,5}Z+d, and 0.02 kcal/mol with V{Q,5}Z+d, provided either the A{4,5} from RP and Schwenke, or the presently optimized version, are used: simple $L^{-3}$ extrapolation triples or quadruples the error, bringing it in the range of *the next basis set pair down* with optimized exponents. At any rate, considering that reaching 0.1 kcal/mol RMSD in valence CCSD/{5,6} calculations will be the practical limit[80] for the CCSD term, it seems quite justified to limit the expensive and memory-hungry (T) calculation to {4,5} basis sets.

For the still more economical {3,4} pair, 0.05 kcal/mol RMSD or better is achievable; $L^{-3}$ is basically equivalent to the optimum here, as can be seen from the α exponents corresponding to our various optimized A{3,4}. For the Weigend-Ahlrichs[99] def2-TZVPP and def2-QZVPP basis sets, α=3.155 is optimum, but the error from using just $L^{-3}$ is quite tolerable. The same goes for {3,4}ZaPa.[4]

In response to a suggestion by Prof. John F. Stanton (U. of Florida), we attempted 3-point extrapolation using $E_\infty + AL^{-3} + BL^{-5}$. With the help of the Mathematica computer algebra software, the closed-form solutions for L={2,3,4}, {3,4,5}, and {4,5,6}, respectively, are found to be:

$$E_\infty\{2,3,4\} = E[Q] + (673\,E[Q] - 729\,E[T] + 56\,E[D])\,/\,607 \qquad (26)$$
$$E_\infty\{3,4,5\} = E[5] + (14197\,E[5] - 16384\,E[Q] + 2187\,E[T])\,/\,7678 \qquad (27)$$
$$E_\infty\{4,5,6\} = E[6] + (12809\,E[6] - 15625\,E[5] + 2816\,E[Q])\,/\,4687 \qquad (28)$$

However, as seen in footnote (g) of Table 3, these offer no advantage over two-point extrapolation with fitted exponents.

Finally, dropping down to the least expensive {2,3} pair entails 0.2 kcal/mol RMSD or worse, except when using the cc-pV{D,T}Z-F12 basis sets for explicitly correlated calculations in an orbital context. Interestingly, the optimum A{2,3} for the A'V{D,T}Z+d pair corresponds to α=3.155, not too far from the α=3.22 obtained from a small training set in W1 theory[76] (which yields essentially the same RMSD=0.19 kcal/mol as the optimum).

What about (T) in CCSD(T)-F12b calculations? First of all, while F12 geminals[101] and explicitly correlated methods[78, 79] more generally, greatly accelerate basis set convergence of the CCSD correlation energy, Köhn[102, 103] showed in great detail that they do not benefit (T) in any way. Second, we find here that for (T), the V{Q,5}Z-F12 basis set pair does not seem to

---

offer any advantages over V{T,Q}Z-F12, neither for CCSD(T)-F12b nor for orbital-only CCSD(T). Third, (T) corrections from CCSD(T)-F12b actually seems to be *inferior* in quality to those obtained from similar basis sets with ordinary CCSD(T) (which will have more modest scratch storage requirements). Fourth, an error below 0.05 kcal/mol RMS is quite achievable using the {T,Q} pair, and 0.1 kcal/mol using the {D,T} pair.

We shall now compare performance for the whole W4-17 dataset. Based on its small RMSD=0.004 kcal/mol from the reference for W4-08, we can use A'V{5,6}Z+d as the 'secondary standard'. Clearly, A'V{Q,5}Z+d at RMSD=0.01 kcal/mol is the basis set pair of choice for high-accuracy work, but V{T,Q}Z-F12 can easily meet a ±0.05 kcal/mol target.

Considering the fairly constant $E_{4T}/E_2$ ratios, the mind wonders if these cannot be exploited to yield a parameter-free estimate for (T) at the basis set limit according to the equation:

$$E(T)[CBS] \approx E(T)[basis] \text{ x } E_{corr}CCSD[CBS]/E_{corr}CCSD[basis] \qquad (27)$$

This is actually a generalization of Marchetti-Werner scaling[147] which we previously considered in Ref.[137]. We can substitute here CCSD/{5,6}ZaPha for $E_{corr}$CCSD. Then we find from Eq. (1) E(T) for nZaPha (n=3–6) with RMSD=0.123, 0.062, 0.037, and 0.025 kcal/mol, respectively — markedly better than the raw results but clearly inferior to extrapolation.

While the 10ZaP basis set is only available for neon, we could trivially truncate 7ZaPa for other elements to generate 'partial wave' basis sets 7ZaPa-$L_{max}= \ell$ for any $\ell<7$. Statistics for the W4-08 dataset with such basis sets can be found in Table 2. It can clearly be seen there that for small $\ell$, 7ZaPa-$L_{max}= \ell$ basis sets are markedly superior to $\ell$ZaPa basis sets in terms of (T) recovery: for the first few $\ell$, 7ZaPa-Lmax= $\ell$ has an RMSD comparable to ($\ell$+1)ZaPa. In fact, we can recover (T) to better than 0.1 kcal/mol by 7ZaPa-$L_{max}$={2,3} extrapolation, and to better than 0.02 kcal/mol by 7ZaPa-$L_{max}$={3,4} extrapolation. While such basis sets are unwieldy for practical use, the results clearly indicate that radial saturation of the available angular momenta is more beneficial for recovering (T) than adding more angular momenta. This situation is very different from the long-standing received wisdom for the singles and doubles correlation energy, as reflected in the 'principal expansion' structure of both atomic natural orbital[148, 149] and correlation consistent[87, 91] basis sets.

## 4. Conclusions

We can state with confidence that achieving basis set convergence for the valence triple excitations energy is much easier than for the valence singles and doubles correlation energy. In fact, it is quite possible to reach convergence on the order of 0.01 kcal/mol with just QZ and 5Z basis sets, and on the order of 0.05 kcal/mol with just TZ and QZ basis sets.

It appears that radial flexibility in the basis set is more important here than adding angular momenta: apparently, replacing *n*ZaPa basis sets with truncations of 7ZaPa at *L*=*n* gains about one angular momentum for small values of n.

As already noted numerous times by Helgaker et al.[119], by RP, and by the present author and coworkers, basis set convergence of the (T) component is fundamentally different from the CCSD correlation energies, and since partitioning between SCF, valence CCSD, and valence (T) energy components is trivially obtained,[5] there is no reason not to extrapolate these components separately.

Moreover, since basis set convergence for the CCSD component may be drastically speeded up through F12 methods, this allows one to keep basis sets down to at most QZ size.

## 5. Future outlook

Quantum mechanical simulation is clearly here to stay. But where will it go from here?

Advances continue to be made in the area of density functional theory. For molecules, empirical double-hybrid density functionals([150] and references therein) offer a fairly low-cost approach (especially if the RI-MP2[151, 152] approximation is made) that approaches the accuracy of composite wavefunction methods.

Double hybrids for solids have recently been implemented,[153] but intrinsically cannot be applied to conductors or semiconductors because of a denominator singularity in the GLPT2 part. Range-separated hybrids[154] and tuned range separated hybrids[155–157] are alternatives.

For wavefunction calculations, the combination of localized orbital methods[36–39] (which make size scaling of the calculation almost linear) and explicitly correlated approaches[46, 78–80] (which drastically speed up basis set convergence) has recently emerged as a powerful alternative[38, 158–160]. Thus, a rigorous, purely WFT-based alternative to DFT exists for large molecules; as the (T) correction does not benefit from explicit geminal correlation, the observations in the present article on the basis set convergence of (T) will be relevant.

A caveat should be voiced, however, about possible errors from localized approaches in extended systems with significant near-degeneracy correlation (a.k.a., nondynamical correlation, static correlation): in Ref.[161] we have investigated this issue for the structures and transition states of polypyrrols, and concluded that (a) PNO-LCCSD(T) and DLPNO-CCSD(T) are vulnerable to static correlation; (b) the LNO-CCSD(T) approach[39] as implemented in MRCC[31] is much more resilient.

For small molecules, canonical CCSD(T) combined with CCSD(F12*)[162] and with general post-CCSD(T) approaches such as CCSDT(Q)[34, 163, 164] as implemented in the MRCC[31] and CFOUR[165] program systems, offer a pathway to sub-kcal/mol accuracy, as we have shown at length in the W4-F12 paper[46].

But the impact of accurate calculations on smaller systems goes further. We have already mentioned their usefulness as "primary standards" for the parametrization of lower-cost empirical methods (as this has already happened, indirectly, via the large and chemically diverse GMTKN55[166] and MGCDB84[167] training sets which mostly compile earlier benchmark calculations). But in addition, Δ-machine learning[40, 168] offers a very attractive alternative that allows improving a low-cost DFT calculation through machine learning on the difference between low-cost and high-accuracy calculated values for a sufficiently large training set. Especially when the training is ad hoc to the problem at hand, this can be a very valuable alternative, for example for acceleration of molecular dynamics on long time scales.

Finally, such Δ-ML values (or those from a lower-cost empirical DFT functional or composite WFT method) could be used as "secondary standards" (in the analytical chemistry sense of the word) for training classical force fields, be they conventional, polarizable[169], or reactive[170].

---

[5] Unlike the partitioning of the CCSD correlation energy in singlet-coupled and triplet-coupled pairs, which is not uniquely defined for open-shell cases.

Acknowledgments This research was funded in part by the Israel Science Foundation (grant 1969/20) and by the Minerva Foundation (grant 2020/05).

Supporting Information Microsoft Excel workbook with the relevant energetics is available at https://doi.org/10.34933/wis.000243